# An India-specific Compartmental Model for Covid-19: Projections and Intervention Strategies by Incorporating Geographical, Infrastructural and Response Heterogeneity

Sanit Gupta[1], Sahil Shah[2], Sumit Chaturvedi[2], Pranav Thakkar[3], Parvinder Solanki[4], Soham Dibyachintan[5], Sandeepan Roy[6], M. B. Sushma[6], Adwait Godbole[2], Noufal Jaseem[4], Pradumn Kumar[4], Sucheta Ravikanti[7], Aritra Das[8], Giridhara R Babu[9], Tarun Bhatnagar[10], Avijit Maji[6], Mithun K. Mitra[4,*,a], and Sai Vinjanampathy[4,11,*,b]

[1] Department of Mechanical Engineering, IIT Bombay.
[2] Department of Computer Science and Engineering, IIT Bombay.
[3] Department of Aerospace Engineering, IIT Bombay.
[4] Department of Physics, IIT Bombay.
[5] Department of Chemical Engineering, IIT Bombay.
[6] Department of Civil Engineering, IIT Bombay.
[7] Department of Electrical Engineering, IIT Bombay.
[8] Epidemiology and Outcome Research, Real World Solutions, IQVIA.
[9] Indian Institute of Public Health-Bengaluru, Public Health Foundation of India.
[10] ICMR - National Institute of Epidemiology, Chennai.
[11] Centre for Quantum Technologies, National University of Singapore.

* Corresponding authors:
   (a) mithun@phy.iitb.ac.in
   (b) sai@phy.iitb.ac.in

**Abstract**

We present a compartmental meta-population model for the spread of Covid-19 in India. Our model simulates populations at a district or state level using an epidemiological model that is appropriate to Covid-19. Different districts are connected by a transportation matrix developed using available census data. We introduce uncertainties in the testing rates into the model that takes into account the disparate responses of the different states to the epidemic and also factors in the state of the public healthcare system. Our model allows us to generate qualitative projections of Covid-19 spread in India, and further allows us to investigate the effects of different proposed interventions. By building in heterogeneity at geographical and infrastructural levels and in local responses, our model aims to capture some of the complexity of epidemiological modeling appropriate to a diverse country such as India.

# Introduction

Epidemiological compartmental models of the Susceptible-Infected-Recovered (SIR) category and its generalizations [1-7] have proven to be an invaluable tool in modeling the population-level spread of infectious diseases. These models can often serve to study various prevention, mitigation and preparedness strategies - and can help inform the public health response to the pandemic.

The limitations of such modelling often arise from the underlying set of assumptions. One common issue in theoretical models is the inherent uncertainty of the datasets from which model predictions are derived. This issue is amplified in the Indian context by economic constraints, cultural norms and other factors that might delay or avoid seeking care. It is in this context that we construct an India-centric model to understand the qualitative features of the spread of COVID-19 and the impact of intervention policies. Our model, summarised in the section below, is a generalized SEIR model with additional compartments representing pre-symptomatic and asymptomatic populations in addition to symptomatic infected individuals. In order to explicitly model uncertainties in testing rates, we introduce a new compartment which accounts for infected populations who have been tested and found to be positive. This, in conjunction with estimates of true infection numbers, allows us to forecast the spread of the epidemic with a greater degree of accuracy. Furthermore, we incorporate methods from estimation theory to mitigate the uncertainty in the initial data and the uncertainty in the fitted parameters of the model.

One of the major drawbacks of the SIR class of models lies in the assumption of well-mixed populations. In the context of a large and heterogeneous country such as India, such assumptions are clearly not valid at a national scale. Our meta-population model assumes a well-mixed population at a district or state scale and then uses a novel technique to construct a district/state level transportation matrix that connects different districts or states. The transportation matrix is constructed using available Census data [53] and allows for the generation of district-wide predictions for the entire country. These models can be used in conjunction with agent-based models at a finer scale such as ward or city level modeling in order to build a comprehensive picture of disease spread.

Compartmental models can help project the qualitative features of epidemic spread. We note that the quantitative projections depend on the choice of parameters, and hence should be interpreted with caution. The value of this model lies in understanding the effect of different proposed interventions and how local heterogeneous responses to the epidemic, realistic measures of transport connectivity, and infrastructural and healthcare status of different states, can affect the spread of the disease at a national scale. In the rest of this section, we briefly summarize important related work related to compartmental models.

Extended SIR models of COVID-19 disease transmission in different countries in the world including India have been studied previously [16-48]. In [16], a modified SIR model with a new additional compartment that quantifies the symptomatic quarantined infected people is considered to study the COVID-19 outbreak in Mainland China. The references [17] and [18] study the impact of COVID-19 in Australia and India respectively by an age-stratified transmission model. The role of the health care system and clinical capacity in reducing the COVID-19 morbidity and mortality rates are analyzed in [17]. The authors of [18] explore an extended SEIR model to study the effect of mitigation strategies such as social distancing and testing-quarantine and show that the testing-quarantine strategy yields better results compared to the implementation of social distancing alone, and also the testing-quarantine strategy is more sustainable than a prolonged lockdown. As there are a large number of asymptomatic carriers, the authors have introduced a separate compartment for them. However, their model assumes instantaneous implementation of the lockdown which is generally not practical. In our paper, we have incorporated a more realistic scenario, a gradual implementation of the containment or lockdown.

## Summary of Model

The basic model structure is summarized in Fig. 1(a). One of the major difficulties in the detection and treatment of Covid-19 is that a large proportion of those tested positive seem to be asymptomatic [8-11], with studies claiming up to 50-75% of the population is asymptomatic. However, since the disease has a long incubation period (5.1 days [8-9]), these asymptomatic infected can spread the disease to the susceptible population even when they show no symptoms of the disease [17]. Further, existing studies suggest that some proportion of the infected population may never develop symptoms over the course of progression of the disease. While there is some debate over whether these true asymptomatics can act as carriers for the infection, studies have suggested that they may still act as carriers albeit with lower infectivity [12,23]. In order to model this significant proportion of asymptomatic infected population, we introduce two new compartments - True Asymptomatics (E) who never develop symptoms over the course of progression of the disease; and Pre-symptomatics (A) who are currently asymptomatic but will develop symptoms after some incubation period. The pre-symptomatic population become symptomatic infected (I) with some rate $\sigma$ (where $\sigma$ is the inverse of the mean incubation period, taken in this study to be 5.1 days).

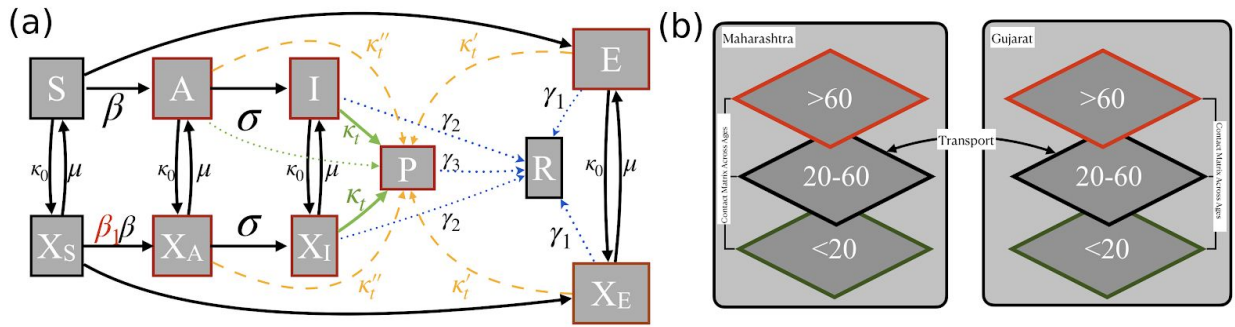

Fig. 1: Panel (a) shows the basic structure of the model. The compartments correspond to the Susceptible (S), Asymptomatic Infected (E), Pre-symptomatic Infected (A), Symptomatic Infected (I), and their corresponding lockdown counterparts ($X_S$, $X_E$, $X_A$, $X_I$). In addition, we introduce a compartment for the population which has been tested positive (P). People can move from the disease compartments (E, A, I, $X_E$, $X_A$, $X_I$, P) to the recovered (R) compartment. The associated rates for each of the transitions are also indicated. Panel (b) shows how this basic building block is replicated for three age groups (<20, 20-60, >60) at each state or district level. States and districts are connected by a transportation matrix for the working population (20-60 age compartment).

One of the major interventions that had been undertaken in India is the imposition of an unprecedented national lockdown since the 26th of March, 2020 to 31st May, 2020 in stages. Following that, there has been a heterogenous lockdown imposed in various parts of the country by state agencies. Any modeling effort to predict, even qualitatively, the future time course of the epidemic in India, must take into account the effect of this lockdown. We introduce shadow compartments for the population in lockdown corresponding to the Susceptible ($X_S$), Asymptomatics ($X_E$), Pre-symptomatics ($X_A$), and the Symptomatic Infected ($X_I$) categories. On the date of imposition of the lockdown, the population starts moving from their normal compartments into the corresponding lockdown compartment with a rate $\kappa_0$. Similarly, on lifting lockdown, people move back from the lockdown compartments to their normal compartment with a rate $\mu$. In this study we have assumed the timescales for this population movement from the normal to the lockdown compartments to be one week in both cases ($\kappa_0$ and $\mu$ are the inverse of this timescale).

We introduce a new compartment for people who have been tested positive for Covid-19 (P). One of the major public health challenges in India (and indeed, worldwide) has been possibly low numbers of people who have been tested, in comparison to projected estimates of "true" infected numbers. Since the accurate estimation of the real magnitude of the infected population critically affects transmission and hence disease progression, we introduce this new P compartment to reflect shortcomings in the testing criteria as well as testing kit availability and the overall healthcare infrastructure. People who have been diagnosed as positive for Covid-19 and hence quarantined in isolation or a medical facility are removed from the general population and hence contain the spread of infection. This population can only infect the healthcare providers, which is incorporated in our model by a term proportional to the fraction of the population who are healthcare workers [12]. The ratio of the number of people who have been tested positive to the "true" number of the infected population gives a measure of the testing fraction $f_{test}$, which then influences the rates at which people travel from the disease compartments (E, A, I, $X_E$, $X_A$, $X_I$) to the tested positive compartment. As such this testing fraction is a critical component of our model as it determines the fraction of the population moving to the P compartment and hence disease progression at a population level. Unfortunately, systematic and accurate estimates of the testing fraction is a complex problem in its own right and is beyond the scope of this paper. Instead we use a combination of statistical inference and empirical methods to arrive at plausible values of the testing fraction for each state. We assume that Kerala, as the state with the best healthcare indicator in the union, has an accurate reporting of Covid-19 related mortality. We then use estimates of the mortality to recovery ratio [54] to arrive at a projected value of the testing fraction for Kerala, given by $f_{test}^{KL} = 0.4$, which implies that Kerala identifies 1 in every 2.5 infected people. We then assign a testing fraction to the individual states by benchmarking their testing performance and healthcare index from the National Health Mission to Kerala's according to the empirical formula. The details of this procedure and the implications for the testing numbers of each state are detailed in the Appendix.

In order to account for different social and economic constraints on different age groups, as well as the well documented age-based differential Covid-19 mortality and severity, we consider a coarse age-stratification within our model where this basic model structure is replicated across three age groups - (i) 0-20 years, (ii) 20-60 years, (iii) >60 years. Such a coarse graining reflects three qualitatively different groups, namely young children and school going adults, working age adults and adults who are more likely to be retired and/or who stay at home. The contact matrices are aggregated according to a weighted average using the age distribution of the Indian population. The contact matrices between these three age groups are obtained from literature and sub-divided into contact matrices corresponding to home, school, work, and other interactions [13]. This stratification of contact matrices allows for exploration of different intervention strategies.

The susceptible population is infected when it comes into contact with the population in any of the disease compartments, with an inherent state-specific transmission rate $\beta$. Within the lockdown period, the home contact matrix is unaffected, while the school contact matrix is completely switched off. However transmission between the population in lockdown and population not in lockdown (essential workers and other partially open sectors) continues via the contact matrices corresponding to work and other interactions, albeit by a reduced factor of $\epsilon_1 (0 \leq \epsilon_1 \leq 1)$. Likewise interactions between two individuals (work and other interactions) in lockdown is reduced by a factor of $\epsilon_1^2$. This parameter $\epsilon_1$ models the leakiness of the lockdown, with $\epsilon_1 = 0$ indicating a perfect lockdown, whereas $\epsilon_1 = 1$ indicates a completely leaky lockdown. Thus in lockdown, disease transmission is not completely halted, but rather continues with a reduced contact matrix, which reflects the realities of a complex Indian society.

Infected people can move from the disease compartments either directly to the recovered (and removed) compartment, or they can first move to the tested positive compartment (P) and thereafter to the recovered (R) compartment. We extract mortality projections from the R compartment by using previously determined estimates of age-stratified mortality for Covid-19.

The meta-populations at a district or state level were allowed to come into contact with each other by estimating the number of people commuting across district boundaries. For this, an inter-district transportation matrix representing the expected number of people commuting across the district border was developed. During biological disasters like COVID-19, the home to work commutes would prevail. Though the comprehensive mobility plan for major Indian cities are readily available, transportation planning studies for the rural districts are hard to find. So, to estimate the number of workers commuting across a district boundary, we used the surrogate measures such as the district level aggregated home to work trip length, worker population density and geographic features. The home to work trip length and worker population details were obtained from the 2011 census data [53], and the geographic features of districts were assimilated in GIS from various sources [56] and verified with available information [57]. Suitable traffic analysis zones (TAZ) were developed by concentric segmentation of the district GIS maps [60]. A trip length frequency distribution developed from the home to work trip length details was used to estimate the trips across the district boundary from each TAZ. Such trips for all the TAZs of a district were added to represent the worker population expected to commute beyond the district boundary. We assumed the work-related commute not to be beyond the adjoining districts and hence, used the relative density of worker population of the adjoining districts to estimate the number of workers commuting to that district. For this estimation, the GDP (at purchasing power parity) per capita of the districts is a better choice of parameter, but was not readily available for all the Indian districts at the time of this study. A detailed description of the process is available in the Appendix E.

There are two "measurements" that must be matched with our model to make predictions about the future of the disease progression. The first of these is the rate associated with mortality, which happens on average about 17 days [14] after the onset of the disease and the second is the number of people that test positive on any given day. Since these two measurements are inherently unreliable, owing to (among other things) socio-economic reasons which might delay seeking care, the data must be treated as inherently noisy. This noisy signal would then lead to variations in the projections of the model. We incorporate this into our model by using an extended Kalman filter (EKF) [15] to estimate the future trajectory by extending the simulation. The EKF is a Bayesian update method that incorporates the measurement errors and allows us to make predictions consistent with the uncertainty in the data. We used the age dependent mortality data available in literature to model the mortality predictions used in the EKF.

## Results

**State-level simulations:**

While our model can simulate a variety of intervention strategies, we propose and investigate two intervention strategies in addition to the base (control) scenario for India. We detail these intervention strategies below:
- Base scenario: We used the initial date for the national level lockdown in India which was the 3rd of May, 2020 as the baseline scenario. As a control simulation, we assumed that lockdown ends on this date, and subsequent to this period, all activities return to normal throughout the nation. The predictions for this scenario can then be used to assess the effectiveness of different proposed interventions.

- Intervention 1 - Enhanced testing: In this scenario, the national lockdown ends on the 3rd of May, 2020. Preparatory to this, all states undertake efforts to enhance testing capabilities and criteria on an emergency basis. We divide states into three categories based on their Health index score by the National Health Mission. This supplies a ranking of the states based on their health care preparedness and infrastructure. We then propose a differential target for each state by benchmarking to the testing performance of Kerala, to be achieved by the 3rd of May. The top third of the states achieve Kerala's testing fraction, the middle third achieves 50% of Kerala's testing rate, while the bottom third achieves 25% of Kerala's testing rate. The increase in the testing fraction is assumed to be linear. The details of the categorization of states and hence their proposed target testing rates are provided in the Appendix.
- Intervention 2 - Heterogenous lockdown: In this scenario, the national lockdown ends on the 3rd of May, 2020. Subsequent to this, schools and colleges remain closed until the end of June, and hence their contribution in the age-stratified contact matrix is switched off. The contact matrices corresponding to work and other interactions are switched on at 50% of their base value, corresponding to continued and stringent enhanced physical distancing measures, public health precautions and enhanced work-from-home situations, wherever possible. The transport matrix is switched on at 50% of its base value, again reflecting a reduction in non-essential travel. If despite these measures, the total tested positive numbers in any state crosses 0.01% of the state's population, the state goes into immediate lockdown. Transport between this affected state and others are completely switched off, the contact matrices corresponding to work and other interactions drop to 20% of their base value within the state. These restrictions continue until the end of June, and testing fractions continue to hold at their currently determined levels.

Though the two proposed interventions are simulated assuming only the first lockdown till March 03, 2020, the purpose of studying the results is to ascertain the effect of transportation and heterogeneity in managing the cases of COVID-19 across the country while minimizing economic impact, which disproportionately affects those whose contact matrices cannot be reduced due to economic necessity. Subsequent to this India went through differing lockdown protocols, which were not simulated in this work. This choice is in part due to the ongoing national and state-level interventions that are evolving rather dynamically even at the time of this writing. The predictions of these interventions should thus be interpreted in a qualitative fashion, taking into account the various assumptions and limitations of the model we have detailed.

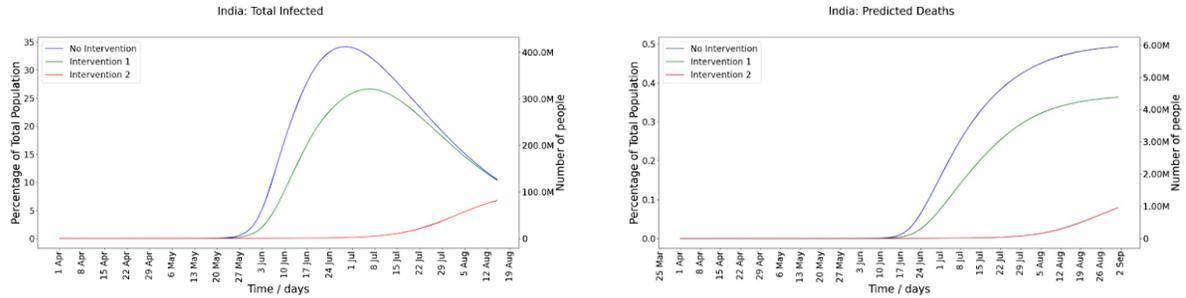

Fig. 2: National level projections of (a) total infected numbers, and (b) predicted mortality estimates from the three scenarios described in the text. Note that the predicted mortality figures show the cumulative numbers. The enhanced testing intervention, where states proportionately improve their testing performance in line with their respective Health index scores provides a measurable improvement in the total infected numbers, and almost a 30-35% improvement in projected mortality. The heterogenous lockdown situation, flattens the curve even further, given that the proposed measures are extremely stringent. Also note that this intervention pushes the peak later in time, beyond the current simulated timeframe.

We first present the results of the base scenario and the two proposed interventions from a state-level national simulation. Figure 2 shows the aggregate national results to provide a comparison of the efficacy of the different interventions. The first intervention - corresponding to enhanced testing with a differential target for each state based on its health index scores - provides a measurable improvement in the total infected numbers, bringing it down from 35% at peak to around 25% at peak. The cumulative number of projected deaths drops from around 6 million to around 4.2 million. While testing at higher rates would yield even better results, for example if all states attempted to measure up to the Kerala rates, this intervention is proposed as an achievable target given the current state of the healthcare infrastructure. The heterogeneous intervention - with reduced contacts and reduced transportation - achieves a major two-fold improvement. On the one hand, it brings down the total number of infected numbers and the total mortality projections because of stringent restrictions. At the same time, it shifts the peak of the curve significantly to the right, delaying the onset of peak, and thus allows the healthcare infrastructure of the country time to set in place better testing and other health infrastructure to manage the pandemic. Under this scenario, the cumulative mortality is at 1 million at the end of August, an almost six-fold improvement on the no-intervention scenario. Note however, that the peak of the infection curve is yet to manifest at this point, which implies that mortality would continue to rise in the absence of any other medical or social interventions.

We now show results from these interventions at the level of an individual state. We show results from six states, chosen as illustrative examples. These states are - Kerala, Maharashtra, West Bengal, Karnataka, NCT of Delhi and Uttar Pradesh. The results for the total infected numbers, the number of diagnosed individuals, and the mortality projections for these states are shown in Figs. 3-8. The corresponding curves for other states are presented in the appendix. The corresponding health indices and testing rates for these six states are - Kerala (70.88; 0.4), Maharashtra (60.58; 0.298), West Bengal (58.06; 0.026), Karnataka (59.22; 0.144), NCT of Delhi (49.04; 0.594), Uttar Pradesh (30.92; 0.036).

These statistics can help interpret the differential effects of the different states under the proposed interventions.

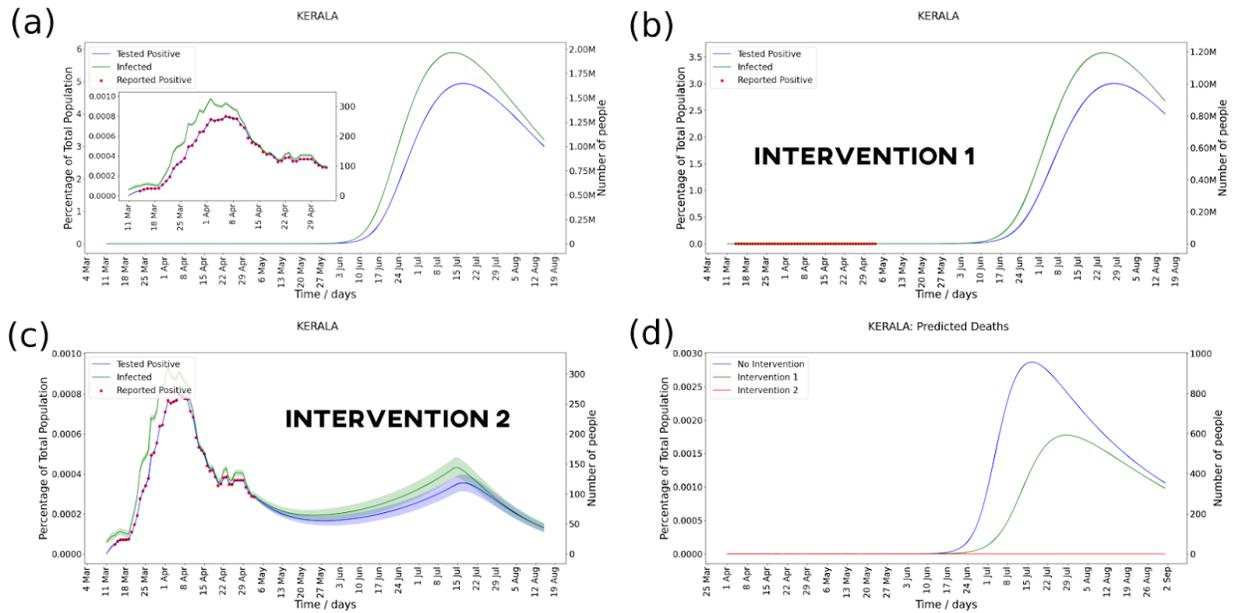

Fig. 3: Comparison of projected infected and mortality numbers for **Kerala** under different proposed scenarios. Panel (a) shows the infected numbers no-intervention scenario, while panel (b) and panel (c) shows the results for interventions 1 and 2 respectively. A comparative analysis of the projected mortality figures for the three situations is shown in panel (d). The inset in panel (a) shows a zoomed in view of the reported number of cases till 3rd May (red dots).

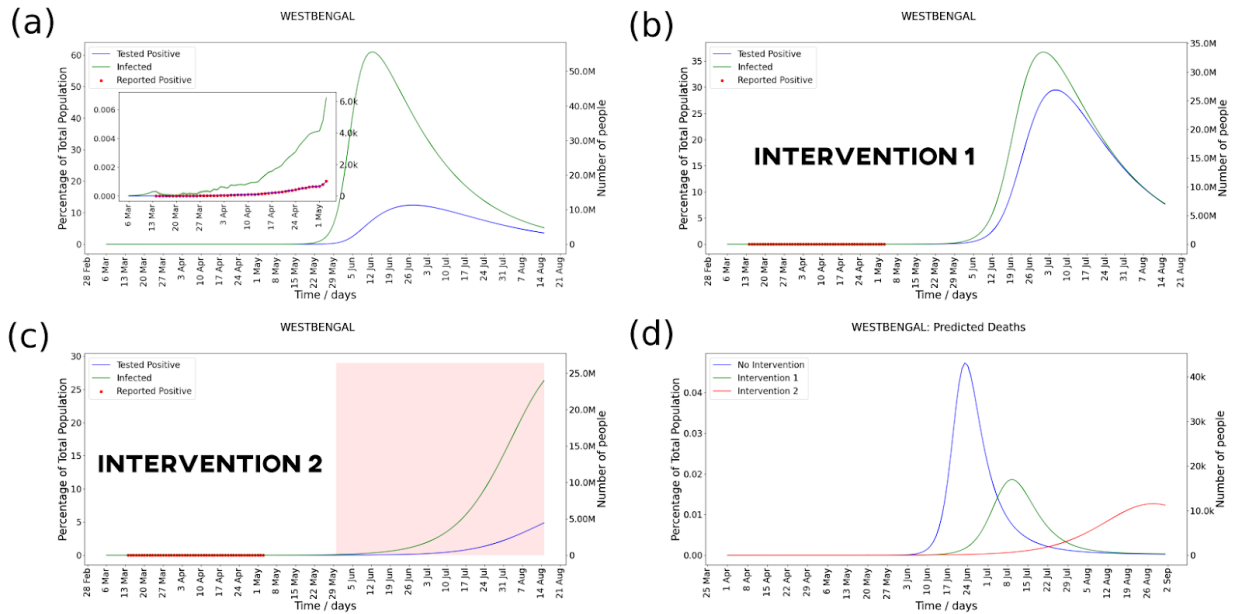

Fig 4: Comparison of projected infected and mortality numbers for **West Bengal** under different proposed scenarios. Panel (a) shows the infected numbers no-intervention scenario, while panel (b) and panel (c) shows the results for interventions 1 and 2 respectively. A comparative analysis of the projected mortality figures for the three situations is shown in panel (d). The inset in panel (a) shows a zoomed in view of the reported number of cases till 3rd May (red dots).

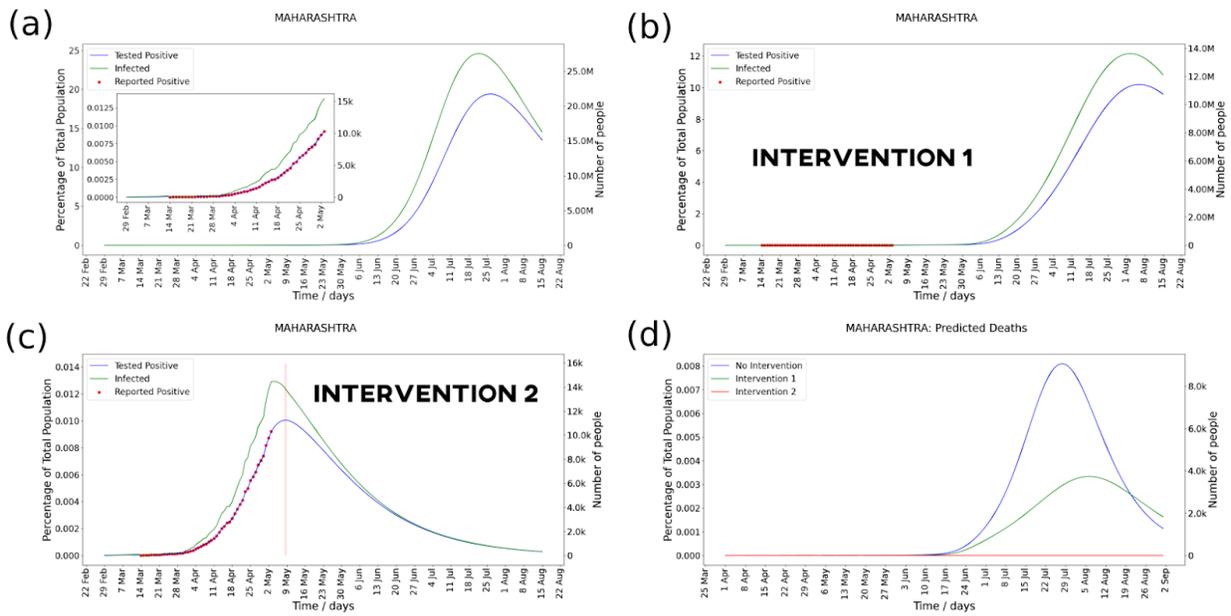

Fig 5: Comparison of projected infected and mortality numbers for **Maharashtra** under different proposed scenarios. Panel (a) shows the infected numbers no-intervention scenario, while panel (b) and panel (c) shows the results for interventions 1 and 2 respectively. A comparative analysis of the projected mortality figures for the three situations is shown in panel (d). The inset in panel (a) shows a zoomed in view of the reported number of cases till 3rd May (red dots).

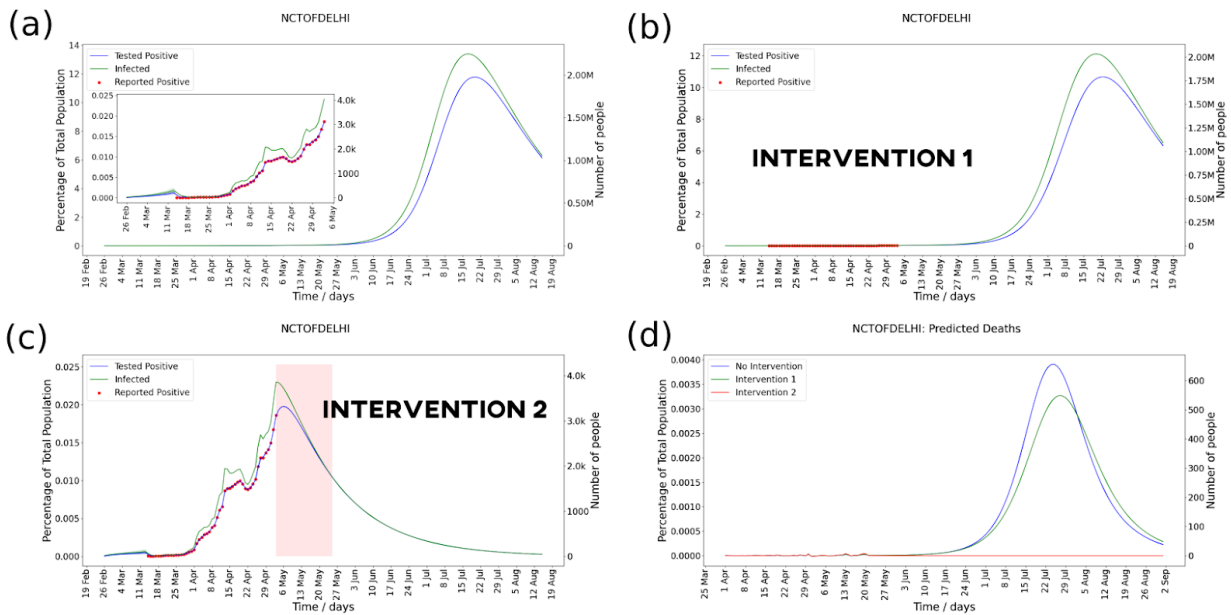

Fig 6: Comparison of projected infected and mortality numbers for the **National Capital Territory of Delhi** under different proposed scenarios. Panel (a) shows the infected numbers no-intervention scenario, while panel (b) and panel (c) shows the results for interventions 1 and 2 respectively. A comparative analysis of the projected mortality figures for the three situations is shown in panel (d). The inset in panel (a) shows a zoomed in view of the reported number of cases till 3rd May (red dots).

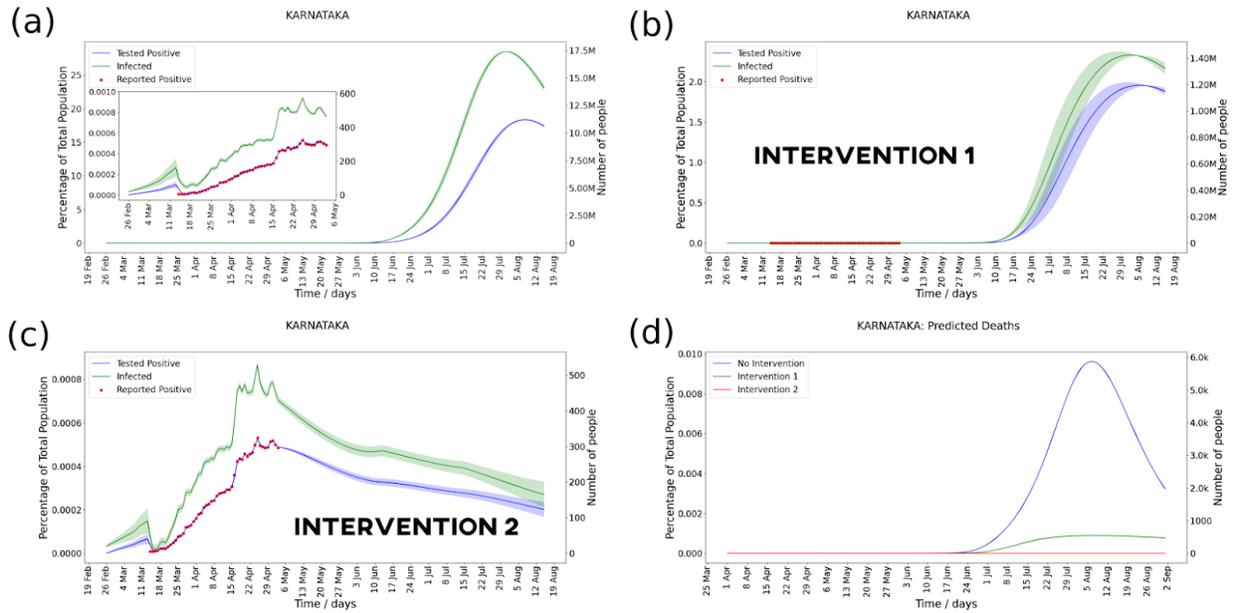

Fig 7: Comparison of projected infected and mortality numbers for **Karnataka** under different proposed scenarios. Panel (a) shows the infected numbers no-intervention scenario, while panel (b) and panel (c) shows the results for interventions 1 and 2 respectively. A comparative analysis of the projected mortality figures for the three situations is shown in panel (d). The inset in panel (a) shows a zoomed in view of the reported number of cases till 3rd May (red dots).

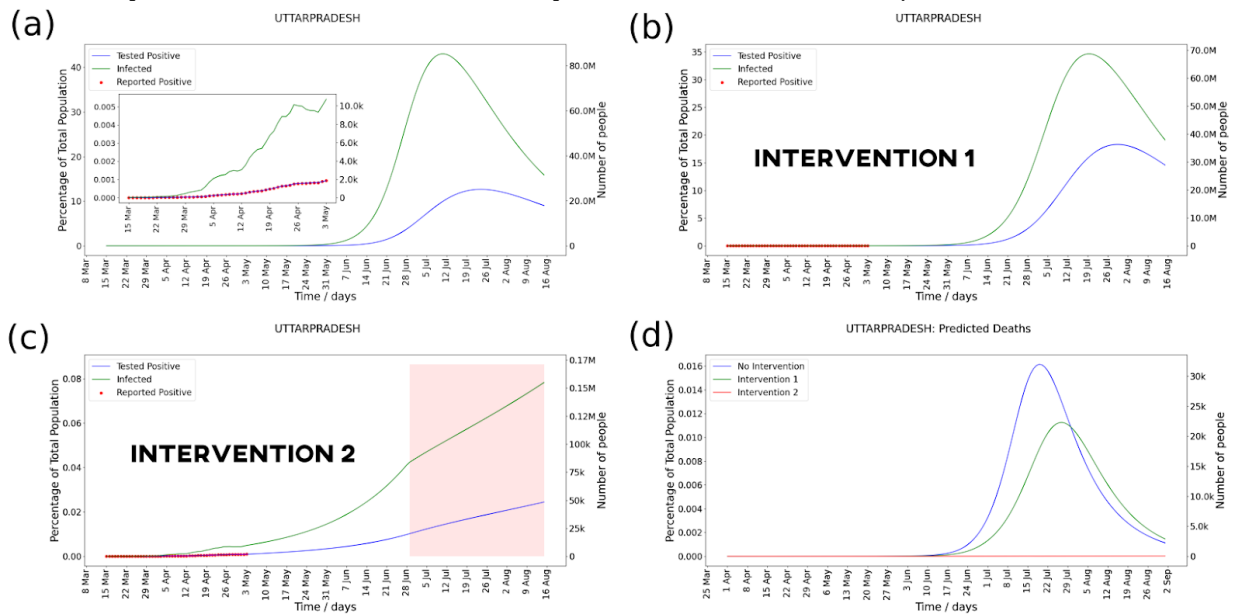

Fig 8: Comparison of projected infected and mortality numbers for **Uttar Pradesh** under different proposed scenarios. Panel (a) shows the infected numbers no-intervention scenario, while panel (b) and panel (c) shows the results for interventions 1 and 2 respectively. A comparative analysis of the projected mortality figures for the three situations is shown in panel (d). The inset in panel (a) shows a zoomed in view of the reported number of cases till 3rd May (red dots).

First, we discuss the baseline no-intervention scenario. Kerala has some of India's best healthcare and socioeconomic indicators alongside a robust public healthcare system. The large testing rates observed thus far in Kerala are reflected in the large testing fraction chosen in our model. This causes all three scenarios to have modest increases in disease progression (in % population) in comparison with other states. For instance, in the no-intervention scenario, the peak infected numbers are about 6%. This is contrasted against the no-intervention scenarios of West Bengal, which peaks at 60% (note also the much larger population of West Bengal). Likewise, states such as NCT Delhi show more favourable disease progressions than neighbors such as Uttar Pradesh which have lower healthcare indices and fewer tests per million.

To compare interventions, we will use the peak simulated infection as a percentage of the population of the states. Since India has on average about 0.50 beds per thousand people [59], this is a good measure of the potential load on the healthcare system. This is important given that cities such as Mumbai and Delhi have already seen a scarcity in available hospital beds. Other measures of interest such as total infecteds and total mortality track the peak infected percentage and can be inferred from the figures in the main text and appendix. Both our proposed interventions do well in lowering simulated infection rates and lowering the infections, though their effectiveness strongly depends on the state. For instance, under intervention 1, Kerala's peak infection rate sees a reduction from 6% to approximately 3.5% whereas West Bengal sees a reduction from 60% to 35%. Gujarat's peak infections reduce from 40% to approximately 25%, once again, an overall reduction by a factor of 1.6. Karnataka on the other hand shows a dramatic improvement from 26% to 2%, which is a reduction in peak infections by a factor of 13. To understand this, we look at the testing rates and health care indices of these states. Kerala has one of the highest testing rates in the country at 0.4 in comparison with Gujarat which has a testing rate of 0.235, and both states are in the upper tertile of the health index rankings. In contrast, Karnataka has a relatively high health index ranking putting it in the upper tertile as well, though its current testing rate is quite low at approximately 0.144. Under intervention 1, Karnataka's testing rate was moved up to Kerala's value at 0.4. We believe this alongside the reduction in contact matrices explains the rapid reduction seen in Karnataka's case. To further inspect this, we look at two other states with proximate testing rates, namely Punjab and Himachal Pradesh. Himachal Pradesh has a relatively high health index score, and hence moves to Kerala's testing rate, causing the peak infection to come down from 12% to approximately 2%. In contrast to this, Punjab has a high Health index putting it in the upper tertile, and hence its simulated results should resemble Karnataka and Himachal Pradesh in showing large reductions in infection rates. We only see approximately a 2 fold decrease in the disease progression as measured by the aforementioned ratio. We believe that this discrepancy might be due to differences in the transport rate between these states.

In comparison to the baseline scenario and intervention 1, our proposed intervention 2 is far more severe, where a heterogenous lockdown is imposed on a strict threshold set at 0.01% of the state's population. This period of heterogeneous lockdown is represented in the figures as the shaded region.

Such a serious intervention clearly produces drastic results, reducing the simulated peak infections to a manageable 0.02% population for Gujarat and 0.001% of the population in Kerala. Likewise several other states show a dramatic reduction in the number of infections under intervention 2. The state of West Bengal is an exception to this trend, where under the strict intervention 2 there is no abetment of the spread of the infections. This is likely due to the fact that the low (existing) testing rate of West Bengal is insufficient to curb the growth of the infections. This yet again highlights the importance of testing.

**District level simulations:**

A major feature of our model is that capability to generate district level predictions for the entire country. As a demonstration of this level of granularity, we simulate the time course of the pandemic at a district level for the whole of India until the end of May. The data for the reported cases, recovery and mortality data was obtained till the 3rd of May, and the simulations were run until 26th of May assuming that lockdown was lifted post 3rd May, and normal life and transportation resumed to pre-lockdown levels. The results are shown in Fig. 9. Note that, as with all the results of the model, these predictions depend on the various model assumptions and parameter values, and should only be interpreted in a qualitative sense. Note that these projections are not meant to simulate reality since the nation went through various levels of lockdown or varying heterogenous severity during this period. Our simulation aims to show that national and even state-level predictions can often mask a wide range of performances depending on the level of granularity, and therefore accurate models aiming to predict the pandemic should take this granularity into account. In addition, improved parameter estimates can help provide more accurate predictions. Also, note that for districts which have not reported any cases, one can in principle adopt two approaches to these "silent" districts - one can trust the reporting of the data, and assume there have been no cases in these districts; or one could assume that the reporting itself is uncertain, and the true numbers in these districts are unknown. We show in Fig. 9 the results corresponding to the first assumption, where we trust the data reporting. The colour map corresponds to the total number of pre-symptomatic and symptomatic infected people, as well as all diagnosed cases ( $A + I + X_A + X_I + P$ ).

Note that the country as a whole presents a very heterogeneous status of disease progression at a given point of time. The testing rates introduce an immediate source of variability, while the transportation matrix ensures a disparate spread of the disease once lockdown is lifted, even within districts of the same state. Note that since this full lifting of lockdown and a return to pre-lockdown contact rates is not reflective of the true situation, the model projects numbers which are much higher than the true reported number of infections. A faithful prediction would involve incorporating time-varying heterogenous interventions for different districts, as was done. While our model is capable of incorporating such heterogeneous interventions, we do not choose to attempt to simulate this since the policy for individual states was often unclear. Another simplification we have made in the current model is that the transmission rate is assumed to be the same for all districts within a given state, and was determined by fitting the state data to the model. For accurate projections, the data for each district needs to be fitted individually. We choose not to do this since many districts have very few cases such that an accurate determination of the transmission rate is not feasible.

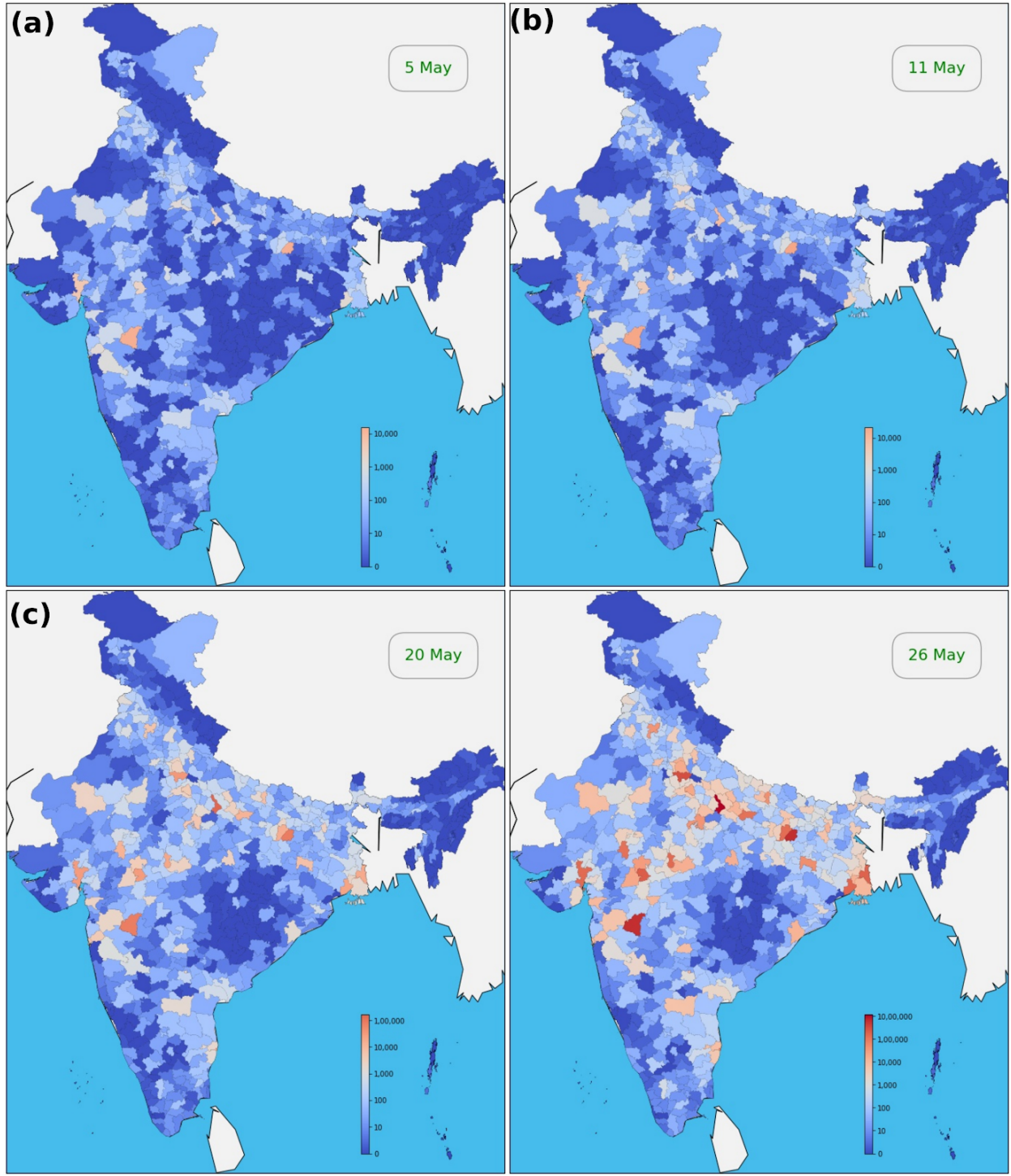

Fig. 9: District level predictions of the projected number of cases at the end of May. The total number of infections are reported as the sum of the Presymptomatic (A), Symptomatic Infected (I), the corresponding lockdown compartments ($X_A$ and $X_I$), and the diagnosed population (P). We do not show the true Asymptomatic population (E and $X_E$) since they never display symptoms and hence would not be a burden on the healthcare infrastructure.

### Effect of transportation

We now turn to quantifying the effect of our mobility matrix on the spread of infection in the country. We simulate the country at a district level using input data until the 3rd of May, and then run our simulations until the 26th of May, under two scenarios - with the transportation matrix switched on, and with the transportation switched off. All other conditions, including resumption of all contacts to pre-lockdown levels were kept constant between the two simulations. The comparative district wide plots on the 5th of May and 26th of May are shown in Fig. 10. For the 5th of May, since this is immediately after lockdown, the differences in the transportation vs no. transportation scenarios is minimal, as can be seen in Fig. 10(a). However, with time, the unequal mixing due to resumed mobility of the population kicks in, leading to drastic differences in infected numbers, shown in Fig. 10(b).

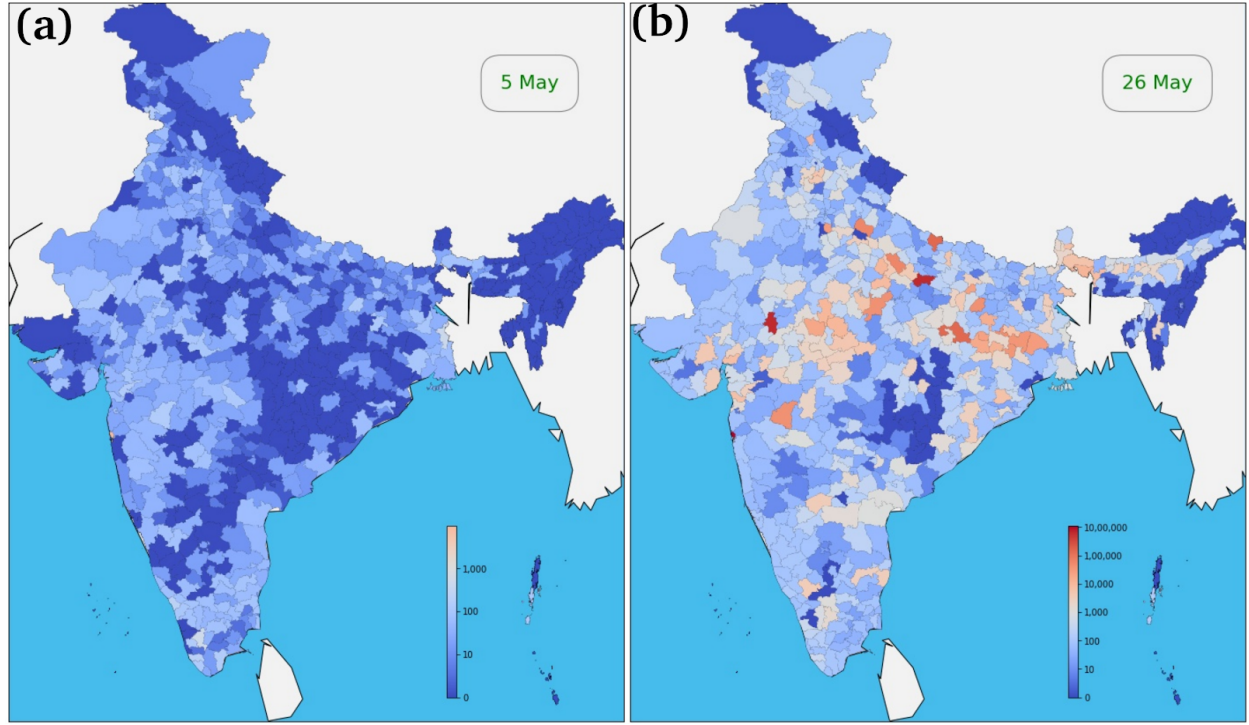

Fig. 10: Difference in infection numbers with and without transportation at a district level. The difference is shown for two dates - (a) May 5th, soon after lifting of lockdown, and (b) May 26th, around 3 weeks after lifting of lockdown. The mixing due to transportation causes major surges in infection numbers in select areas, as can be seen in the 26th May plot.

We wish to highlight two important features that can be attributed to the effect of transportation. Firstly, there is an overall rise in the number of cases due to mixing of the susceptible and infected populations due to transportation - as is expected. More interestingly, there are emergence of clusters with very high infection numbers when the transportation matrix is switched on. Some such sample clusters are marked in the figure. This highlights how transportation can bring in new infections and cause a growth of cases in areas that would otherwise have the epidemic under control. Note that this effect is heterogeneous - there are areas in the country where the infection stays within control even with transportation. This highlights the inherent disparity in transportation patterns, and suggests that there are more effective methods of epidemic control instead of blanket bans of travel. Transportation may be allowed in different areas at different rates while still keeping the spread of the epidemic in check.

## Comparison of predicted and observed trends

Our model aims to provide a framework within which to incorporate multiple levels of heterogeneity appropriate to a country such as India. Accurate quantitative projections require dynamic updating of the model to take into account time variations of parameters such as testing rates, higher quality and more reliable data, and clarity regarding heterogeneous policy implementations at the district and state levels. Nevertheless, in order to benchmark the basic accuracy of our model, we show a comparison of the predicted trends under the base scenario and the two proposed interventions in this manuscript with the actual reported data for the number of diagnosed Covid-19 cases. Note that for our simulations, the reported data for positive population and the Covid-19 associated mortality were incorporated until the 3rd of May, and the simulations were then run until the end of August. The reported data was considered until the 15th of July. The comparison between the different scenarios is shown in Fig. 11. The data qualitatively agrees with the predicted trend under intervention 2, which corresponds to a heterogeneous lockdown scenario with reduced contacts due to stringent physical distancing and public health measures, reduced nationwide transportation, and local lockdowns depending on the severity of the outbreak. We note that this is the scenario which agrees strongly with the actual policies implemented in India post the end of the national lockdown on the 3rd of May. The base scenario considers the situation where post-May 3rd, the nation returned completely to pre-lockdown levels of contacts and transportation. This was evidently not the case, as India went through varying levels of lockdown post May 3rd, and varying levels of unlock protocol as well, post June 8th. The first intervention - again corresponding to resumption of normal life post May 3rd, with the addition of health-performance based enhanced testing, also does not correspond to the real-life situation for similar reasons. Thus it is expected that the base scenario and intervention 1 predicts much higher infection numbers compared to the actual data. The broad qualitative agreement between the predictions of the second intervention with the actual data serves as a strong consistency check on the background assumptions and methodology of the model, even in the absence of implementation of granular district level intervention policies and testing rates. Incorporation of these granular details in the model can be used to generate more accurate quantitative predictions to guide future policy.

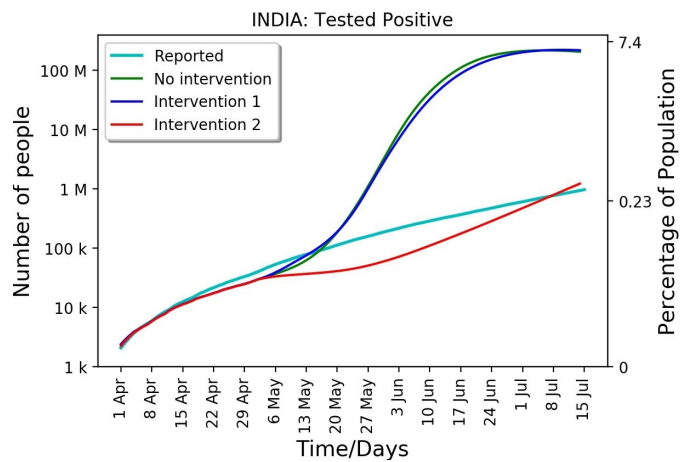

Fig. 11: Comparison of the reported number of positive cases with the three intervention scenarios proposed and studied here. The data is till 15th July, 2020, while the predictions were generated using a dataset till 3rd May, 2020.

**Discussion**

We use a simple age stratified model in order to capture differential disease outcomes, intervention strategies, and transportation effects. We use three age categories - people under 20 years, people between 20 to 60 years and people above 60 years. While a finer stratification is available, we chose these three coarse grained age stratifications in order to minimize the number of compartments, and hence parameters in the model, while at the same time ensuring that one can test different India specific interventions. One common strategy is to ensure closure of schools and colleges across the country for an extended period of time, which affects the 0-20 age bracket. Further, the inter-district transportation rates were computed for the worker population of a particular district, and hence these were incorporated only in the 20-60 year age bracket. Further, projected mortalities were estimated using an age-dependent mortality rate, as is relevant for Covid-19.

An important ingredient in our model is a state-wide identification of testing rates. This is important because it allows us to build in a state-level heterogeneity that incorporates - (i) the Health Index and (ii) the number of covid tests per million population. The Health index provides a measure of the public health infrastructure of states, including hospital bed availability, number of ASHA and ANM personnel and other relevant health indicators. This is important because the state of the public health infrastructure constrains the response of a specific state. Contact tracing and other measures depend on the state of health preparedness of the state and hence is an important metric that must be taken into account. The second metric, of the number of Covid-19 tests depends on the testing policy and on availability of testing kits and also affects identification and quarantine efforts. It should be emphasized that the numbers for the testing rates may not be the "true" numbers, however, this empirical estimate helps capture the heterogeneous response in different states of the country and hence critically affects outcomes and projections at a national level. Finally, we note that the testing rates are taken as (static) numbers in our model, whereas they are perhaps more accurately described as time series. Incorporating the time series as a determinant of the testing rates can help provide more quantitative prediction for the progression of the epidemic. However, we note that there have been consistent concerns about the accuracy of the data reporting, and further, that accurate quantitative projections require granular details of local intervention policies, which are currently not available for the nation as whole. Since our aim was to provide a framework of incorporating heterogeneity in testing rates, we choose to restrict the model to static rates for the purposes of the current analysis.

An important contribution of the model is to estimate and construct a national transportation matrix at a district-level from available census data. Our model simulations show the importance of transportation in the spread of the epidemic and underscores the importance of incorporating this in any theoretical study. Although the transport network in this paper is constructed by inference, as described in the methods section, future work should focus on building real data-driven mobility networks for the entire country at a granular level. We also note that the national level lockdown also set in motion an unprecedented large scale movement of migrant labourers throughout the country. Similar large scale migration also happened with the onset of the unlock protocols. The transportation matrix proposed in this work builds on mobility patterns in normal times and does not take into account this large scale migration. Quantitative estimates of the patterns of this migration can help in improving the model and producing better quantitative predictions.

The limitations of our model should be emphasised to interpret the results in the appropriate context. The first limitation of our model is the fact that compartmental models are built on the assumption that the underlying population is well mixed. Though this is somewhat mitigated by the fact that our model has structure (age structure and geographic specificity), the mixing assumption still implies that we have fluctuations not captured by our model. Furthermore, the aggregation of the contact matrices into three age bins is an approximation to the true dynamics. We also include the health indices of each state into our model by the assumptions detailed in the description of the interventions. The separation of all states into three groups assigned to execute intervention policies is a further approximation that allows us to incorporate the health indices in a straightforward fashion. Simplifying assumptions were also made about the infection rates. The district infection rates are pegged to the state infection rates and the coefficient $\beta_1$ that models the inefficiencies in the lockdown are pegged to the national/state level in the appropriate simulation. Another assumption of our model is that the testing rates are not functions of time.

Besides this, our model is built on simplifying assumptions about modelling various parameters via their median values, though this can be generalized easily. For instance, we match model predictions against reported mortality by assigning 2% of the simulated infections to become fatal in a median of 17 days. Both these numbers are better approximated as probability distributions. These distributions are often long-tailed, and hence a mean parameter value may provide poor projections. A stochastic version of this model which incorporates these underlying probability distributions would be more relevant from a predictive approach. Finally, we made the simplifying assumption that the noise model for the measurement and state errors in the extended Kalman filter are Gaussian as described in the methods section. We presume that this is adequate for large populations but highlight it here for the sake of completeness.

Our model presents a comprehensive compartmental model that is relevant in the context of India. We take into account a detailed generalization of an SEIR model that accurately reflects the disease dynamics of Covid-19. In addition, we use state level health metrics and Covid-19 responses to model the heterogeneous spread, which is essential for a country of the scale of India. Further, our district-level transportation matrix uses Census data to project the number of people coming into and moving out of a particular district and hence builds in a geographical heterogeneity. Our proposed interventions must then be interpreted against the background of this realistic national model. We show how a combination of various strategies may be utilised to emerge from this pandemic while minimizing the human and economic costs of the pandemic.

We hope that our model is relevant in the context of organising a more comprehensive solution to the ongoing pandemic in India. Our model, which is documented with open source code made available online [50] can be modified to qualitatively model both the spread of the disease and the economic cost of interventions. Furthermore, in addition to age and district level stratification our model may be generalized to include income groups. Such a generalised model would potentially address how different socio-economic groups are affected by the pandemic and the onset of the loss of income due to lockdown. It is generally accepted that the optimal strategy to control the pandemic in any region with an active lockdown is to use that time to reinforce testing facilities, hospital beds and personal protective equipment. We hope that our model plays some role in assessing the risks relative to the rewards in combating COVID-19.

In conclusion, we would like to end with a caveat. For a country of the scale and complexity of India, it is impossible for any single model to predict accurately the course of an epidemic. Local responses, super-spreader events, time varying strategies and interventions, and other spatio-temporal fluctuations, all play an important role in the final trajectory of the epidemic. Models such as these and others, can only serve as a guiding tool to policy-makers. However, a holistic approach to the pandemic requires a combination of expertise from public health officials, health-care workers, epidemiologists, sociologists, and economists, in addition to modelers, to come together to determine the most effective response, contingent on the available data. This present work should be taken in the spirit of being only one component in this multi-faceted approach, and should not be considered in isolation.

**Acknowledgements**:

MKM and SV acknowledge discussions with S Gupta and D Manjunath. SV acknowledges a cluster funded by an IITB-IRCC grant number 16IRCCSG019 and a DST-SERB Early Career Research Award (ECR/2018/000957) used for the simulations.

## Supplementary Methods and Materials

### A. Model equations

The meta-population model presented in this manuscript can be simulated at a district or state level. Let, $S^{\alpha i}$ denote the susceptible population in the age group α in the $i^{th}$ district/state. Similarly, $E^{\alpha i}$, $A^{\alpha i}$, $I^{\alpha i}$ denote the true asymptomatic, pre-symptomatic, and symptomatic infected population in the age group α in the $i^{th}$ district/state, while $X_E^{\alpha i}$, $X_A^{\alpha i}$, $X_I^{\alpha i}$ denote their counterparts under lockdown respectively. Finally, $P^{\alpha i}$ denotes the number of people tested positive, and $R^{\alpha i}$ the number of recovered and removed people in the age group α in the $i^{th}$ district/state. Since we use three age brackets, α = 1, 2, 3 denotes the age groups < 20, 20 − 60, and > 60 respectively, while $i = 1, ..., N_{d/s}$, where $N_d = 720$ is the total number of districts in India, and $N_s = 36$ is the total number of states and union territories in India.

The dynamics of the susceptible population is given by,

$$\begin{aligned}
\partial_t S^{\alpha i} &= -\beta_i \sum_{\alpha'=1}^{3} \frac{1}{N^{\alpha' i}} \{C_{\alpha\alpha'}(I^{\alpha' i} + A^{\alpha' i} + b_1 E^{\alpha' i}) \\
&\quad + b_2 \partial_{\alpha,2} P^{\alpha' i} + C_{\alpha\alpha'}^L (X_I^{\alpha' i} + X_A^{\alpha' i} + b_1 X_E^{\alpha' i})\} S^{\alpha i} \\
&\quad + (1 - \Xi(t, T_L, T_E)) \sum_{j=1}^{N_{d/2}} \left( M_\alpha^{ij} \frac{S^{\alpha j}}{N^{\alpha j}} - M_\alpha^{ji} \frac{S^{\alpha i}}{N^{\alpha i}} \right) \\
&\quad - \Xi(t, T_L, T_E) k_0 S^{\alpha i} + \Theta(T_E) \mu X_S^{\alpha i}
\end{aligned}$$

The susceptible population becomes infected on contact with infected individuals (asymptomatic, presymptomatic, and symptomatic) across all age groups. $_i$ denotes the inherent infectivity for the $i^{th}$ district. We assume that the true asymptomatics infect with a lower infectivity ($b_1 < 1$) than presymptomatic or symptomatic individuals. The diagnosed population can only infect healthcare workers and this lower infectivity is captured by the parameter $b_2(\ll 1)$. The contacts between different age groups were decomposed into home, school, work, and "others" and were calculated using the data in Prem et. al. [49]. For the three tiered age structuring in our model, the contact matrices are (3 × 3) matrices and are given by,

$$C_{home} = \begin{bmatrix} 2.007 & 2.326 & 0.153 \\ 1.729 & 1.666 & 0.131 \\ 1.339 & 1.781 & 0.184 \end{bmatrix} \quad C_{school} = \begin{bmatrix} 5.391 & 0.437 & 0.006 \\ 0.743 & 0.380 & 0.006 \\ 0.122 & 0.073 & 0.029 \end{bmatrix}$$

$$C_{work} = \begin{bmatrix} 0.104 & 0.385 & 0.004 \\ 0.205 & 3.793 & 0.059 \\ 0.019 & 0.483 & 0.012 \end{bmatrix} \quad C_{others} = \begin{bmatrix} 6.801 & 3.158 & 0.210 \\ 1.250 & 3.937 & 0.146 \\ 0.180 & 0.916 & 0.209 \end{bmatrix}$$

The total contact matrix is the sum of all four of these constituent matrices,
$C = C_{home} + C_{school} + C_{work} + C_{others}$
Since all lockdowns are imperfect, the susceptible population can come into contact with the infected compartments in lockdown as well. This contact is however reduced from the non-lockdown contact matrix $C$ for work and other contacts by a factor $_1$, which then indicates the leakiness of lockdown. We set the school contact matrix to zero since schools are completely closed during this period. This lockdown contact matrix is then given by,
$C^L = C_{home} + {_1}(C_{work} + C_{others})$.
The second term models the mobility of the working population into and out of a particular district. This transport term is effective only if lockdown is not active, which is implemented by the $(1 - \Xi(t, T_L, T_E))$ prefactor where $\Xi(t, T_L, T_E) = 1$ if $T_L \leq t \leq T_E$ and zero otherwise, and $T_L$ and $T_E$ denotes the start and end dates for the national lockdown. The mobility or transportation matrix $M_\alpha^{ij}$ denotes the number of people in age group $\alpha$ who leave from district $j$ to arrive at district $i$ per day. Our transportation matrix, as detailed subsequently, is calculated for the working age population, and hence $M_1^{ij} = M_3^{ij} = 0$.
The final two terms in the time evolution of the susceptible population equation models the rate $(k_0)$ at which susceptible population locks down. The lockdown is effective from time $T_L$ to $T_E$. After the lockdown is lifted at time $T_E$, the population returns to a susceptible state at a rate $\mu$, where $\Theta$ represents the usual Heaviside function.

Now, of the susceptible population who become infected, we assume that a fraction $f$ are the true asymptomatics (E), i.e. people who will never show any symptoms over the entire progression of the disease. The time evolution of this compartment is given by,

$$\partial_t E^{\alpha i} = +f\beta_i \left[ \sum_{\alpha'=1}^{3} \frac{1}{N^{\alpha' i}} \left\{ C_{\alpha\alpha'} \left( I^{\alpha' i} + A^{\alpha' i} + b_1 E^{\alpha' i} \right) \right. \right.$$
$$\left. \left. + b_2 \delta_{\alpha,2} P^{\alpha' i} + C^L_{\alpha\alpha'} \left( X_I^{\alpha' i} + X_A^{\alpha' i} + b_1 X_E^{\alpha' i} \right) \right\} \right] S^{\alpha i}$$
$$+ (1 - \Xi(t, T_L, T_E)) \left[ \sum_{j=1}^{N_{d/s}} \left( M_\alpha^{ij} \frac{E^{\alpha j}}{N^{\alpha j}} - M_\alpha^{ji} \frac{E^{\alpha i}}{N^{\alpha i}} \right) \right]$$
$$- \Xi(t, T_L, T_E) k_0 E^{\alpha i} + \Theta(T_E) \mu X_E^{\alpha i} - \kappa'_t E^{\alpha i} - \gamma_1 E^{\alpha i}$$

Note that this equation closely mirrors the dynamics of the susceptible population, with a fraction $f$ of the newly infected arriving in the E compartment. The transportation matrix accounts for inter-district/state travel, and the $k_0$ and $\mu$ terms reflect the imposition and lifting of lockdown. The penultimate term models the rate $\kappa'_t$ at which these true asymptomatic people are detected. This can be achieved by contact tracing of known infected individuals. The final term models the recovery of these individuals at a rate $\gamma_1$ which is the inverse of the recovery time for the asymptomatic population.

The remaining fraction $1-f$ of the population transitions into the presymptomatic (A) category. The time evolution of this compartment is given by

$$\partial_t A^{\alpha i} = +(1-f)\beta_i \left[ \sum_{\alpha'=1}^{3} \frac{1}{N^{\alpha' i}} \left\{ C_{\alpha\alpha'} \left( I^{\alpha' i} + A^{\alpha' i} + b_1 E^{\alpha' i} \right) \right. \right.$$
$$\left. \left. + b_2 \delta_{\alpha,2} P^{\alpha' i} + C^L_{\alpha\alpha'} \left( X_I^{\alpha' i} + X_A^{\alpha' i} + b_1 X_E^{\alpha' i} \right) \right\} \right] S^{\alpha i}$$
$$+ (1 - \Xi(t, T_L, T_E)) \left[ \sum_{j=1}^{N_{d/s}} \left( M_\alpha^{ij} \frac{A^{\alpha j}}{N^{\alpha j}} - M_\alpha^{ji} \frac{A^{\alpha i}}{N^{\alpha i}} \right) \right]$$
$$- \Xi(t, T_L, T_E) k_0 A^{\alpha i} + \Theta(T_E) \mu X_A^{\alpha i} - \kappa''_t A^{\alpha i} - \sigma A^{\alpha i}$$

Note the testing of presymptomatic individuals is achieved at a rate of $\kappa''_t$ and these presymptomatic individuals progress to the symptomatic stage at a rate $\sigma$ which is the inverse of the mean incubation period of the disease.

The presymptomatic people develop symptoms and move to the symptomatic infected (I) compartment. The infected population can be diagnosed via positive tests at a rate $\kappa_t$ or they may recover directly without having been diagnosed at a rate $\gamma_2$. In addition they transition to the lockdown compartments and vice versa, and also travel from other districts or states, as in the previous cases. The time evolution of this compartment is then described by,

$$\partial_t I^{\alpha i} = +\sigma A^{\alpha i} + (1 - \Xi(t, T_L, T_E)) \left[ \sum_{j=1}^{N_{d/s}} \left( M_\alpha^{ij} \frac{I^{\alpha j}}{N^{\alpha j}} - M_\alpha^{ji} \frac{I^{\alpha i}}{N^{\alpha i}} \right) \right]$$
$$- \Xi(t, T_L, T_E) k_0 I^{\alpha i} + \Theta(T_E) \mu X_I^{\alpha i} - \kappa_t I^{\alpha i} - \gamma_2 I^{\alpha i}$$

We now turn to the time evolution of the shadow compartments corresponding to the lockdown population. Since lockdowns are leaky, as discussed previously, the susceptible population in lockdown can come into contact with the non-lockdown population with the contact matrix $C^L$ and with other lockdown population with a further reduced contact matrix $C^{LL}$ given by $C^L = C_{home} + \frac{1}{2}(C_{work} + C_{others})$. Note that the lockdown is assumed not to impact home contact matrices at all.

$$\partial_t X_S^{\alpha i} = -\beta_i \left[ \sum_{\alpha'=1}^{3} \frac{1}{N^{\alpha' i}} \left\{ C_{\alpha \alpha'}^L \left( I^{\alpha' i} + A^{\alpha' i} + b_1 E^{\alpha' i} \right) \right.\right.$$
$$\left.\left. + b_2 \delta_{\alpha,2} P^{\alpha' i} + C_{\alpha \alpha'}^{LL} \left( X_I^{\alpha' i} + X_A^{\alpha' i} + b_1 X_E^{\alpha' i} \right) \right\} \right] X_S^{\alpha i}$$
$$+ \Xi(t, T_L, T_E) k_0 S^{\alpha i} - \Theta(T_E) \mu X_S^{\alpha i}$$

Note that for the population under lockdown, the mobility terms do not appear in the time evolution equations, since this population, by definition cannot travel between districts. Analogously, we can write down the evolution equations for the $X_E$, $X_A$, and $X_I$ compartments as follow,

$$\partial_t X_E^{\alpha i} = +f\beta_i \left[ \sum_{\alpha'=1}^{3} \frac{1}{N^{\alpha' i}} \left\{ C_{\alpha \alpha'}^L \left( I^{\alpha' i} + A^{\alpha' i} + b_1 E^{\alpha' i} \right) \right.\right.$$
$$\left.\left. + b_2 \delta_{\alpha,2} P^{\alpha' i} + C_{\alpha \alpha'}^{LL} \left( X_I^{\alpha' i} + X_A^{\alpha' i} + b_1 X_E^{\alpha' i} \right) \right\} \right] X_S^{\alpha i}$$
$$+ \Xi(t, T_L, T_E) k_0 E^{\alpha i} - \Theta(T_E) \mu X_E^{\alpha i} - \kappa'_t X_E^{\alpha i} - \gamma_1 X_E^{\alpha i}$$

$$\partial_t X_A^{\alpha i} = +(1-f)\beta_i \left[ \sum_{\alpha'=1}^{3} \frac{1}{N^{\alpha' i}} \left\{ C_{\alpha\alpha'}^{L} \left( I^{\alpha' i} + A^{\alpha' i} + b_1 E^{\alpha' i} \right) \right. \right.$$
$$\left. \left. + b_2 \delta_{\alpha,2} P^{\alpha' i} + C_{\alpha\alpha'}^{LL} \left( X_I^{\alpha' i} + X_A^{\alpha' i} + b_1 X_E^{\alpha' i} \right) \right\} \right] X_S^{\alpha i}$$
$$+ \Xi(t, T_L, T_E) k_0 A^{\alpha i} - \Theta(T_E) \mu X_A^{\alpha i} - \kappa_t'' X_A^{\alpha i} - \sigma X_A^{\alpha i}$$
$$\partial_t X_I^{\alpha i} = +\sigma X_A^{\alpha i} + \Xi(t, T_L, T_E) k_0 I^{\alpha i} - \Theta(T_E) \mu X_I^{\alpha i} - \kappa_t X_I^{\alpha i} - \gamma_2 X_I^{\alpha i}$$

Finally, we turn to the compartments for those people who have been diagnosed as positive for Covid-19 (P) and the recovered/removed (R) compartment. For the P compartment, the total number of positive cases is the sum of all people who have tested positive from the $E, A, I$ compartments and their lockdown counterparts $X_E, X_A, X_I$. These people recover at a rate given by $\gamma_3$. Thus, we have,

$$\partial_t P^{\alpha i} = \kappa_t (I^{\alpha i} + X_I^{\alpha i}) + \kappa_t' (E^{\alpha i} + X_E^{\alpha i}) + \kappa_t'' (A^{\alpha i} + X_A^{\alpha i}) - \gamma_3 P^{\alpha i}$$

People can recover either from this diagnosed compartment (P) or directly from the symptomatic and true asymptomatic compartments ($I, X_I, E, X_E$) without having been detected over the course of progression of the disease. This recovery compartment then evolves according to the equation,

$$\partial_t R^{\alpha i} = \gamma_1 (E^{\alpha i} + X_E^{\alpha i}) + \gamma_2 (I^{\alpha i} + X_I^{\alpha i}) + \gamma_3 P^{\alpha i}$$

This set of equations completely define the time evolution of the model. At each meta-population level (district or state), there are 30 compartments - 10 population compartments, times 3 age stratifications for each of them. In subsequent appendices, we will discuss the choice of parameters for which this model was simulated.

## B. Parameter estimation

The model presented here is necessarily complicated, given the complex disease trajectory of SARS-CoV-2 as well as the complex societal and geographical realities of India. The parameters of the model sensitively determine the disease trajectory and the geographical spread of the disease. We have used accepted parameters for the disease progression from the literature, while India-specific parameters have been estimated by a variety of methods as detailed below.

Note that different parameter choices will lead to different projections from the model. Hence the numbers reported in this analysis should be interpreted as qualitative trends only, in order to gauge effectiveness of different interventions and the qualitative projected severity of the disease. This model will be continuously updated as and when better estimates of the parameters become available.

| | | Disease progression parameters | |
|---|---|---|---|
| | | Fit | The basic infectivity parameter. This was determined individually for each state by fitting the available data. The full list of the parameter for each state is listed in the subsequent Table |
| $\sigma$ | | 1/5 | The rate of progression from pre-symptomatic to symptomatic individuals. This is the inverse of the mean incubation time of the disease and is taken to be 5 days. [8,9] |
| $b_1$ | | 0.1 | Reduced infectivity of the true asymptomatic population. This is taken to be 10% of the basic infectivity |
| $f$ | | 0.2 | Fraction of true asymptomatics. Assumed to be 20% of the total infected population [10] |
| $\phi$ | | 0.5 | Fraction of population who are asymptomatic at the time of testing. Assumed to be 50% of the total diagnosed individuals [23] |
| $\gamma_1$ | | 1/19 | Recovery rate of true asymptomatics. This is the inverse of the typical recovery time for this population. Since accurate data is not available for true asymptomatics, we use the mean recovery time for mild infections in this case [14] |
| $\gamma_2$ | | 1/22 | Recovery rate of symptomatic infected. This is the inverse of the recovery time post onset of symptoms. This is assumed to be 22 days [51,52] |
| $\gamma_3$ | | 1/22 | Recovery rate of diagnosed population. While the diagnosed or tested positive compartment has influx from different stages of the disease, we use an average recovery time equal to that from the I compartment, since they comprise the largest fraction of people who are tested. |
| | | India specific parameters | |
| $\ell_1$ | | Fit | The leakiness of lockdown. Indicates the reduction in the |

|   |   |   |
|---|---|---|
|   |   | workplace and others contact matrices under lockdown. This was determined individually for each state by fitting the available data. The full list of the $\beta_1$ parameter for each state is listed in the subsequent Table. |
| $b_2$ | 0.002 | The confirmed positive people (P) can only infect health-care workers, since they will be in strict quaran-tine/hospitalization. As a proxy for this parameter then, we can use the estimate of the number of healthcare work-ers in India, which is 20 per 10000 population [12] |
| $k_0$ | 1/7 | This represents the rate at which the general population goes into lockdown. In the absence of concrete data, we assume a timescale of one week. Note that different values of this rate will not affect the disease progression, but will only shift the projected curves. [See A.Dhar [18]] |
| $\mu$ | 1/7 | This represents the rate at which the population exits lockdown. Again in the absence of data, we assume a timescale of one week. |
| $r_{test}$ | Estimated | The testing rate for each state is estimated using a semi-empirical method. We detail the calculation of this testing rate below. The testing rate was then used to obtain estimates of $\beta_t$, $\beta'_t$ and $\beta''_t$. |

**Testing rates:**

In order to compute the testing rates, we follow the procedure outlined in Ref. [58]. The case fatality ratio is defined as $CFR = \frac{\text{Number of deaths}}{\text{Number of deaths} + \text{Number of recovered}}$. Let, D(t) denote the number of deaths at time t, and, r(t) denote the number of recovered people at time t Let us define the ratio, $\rho = \frac{D(t)}{r(t)}$. The CFR can then be estimated from this ratio as $CFR = \frac{\rho}{1+\rho}$. However, because of under-reporting and testing criteria, if the true number of infected I(t) is uncertain, then the true number of recoveries is also uncertain. An analysis for various countries shows that $\rho$ is likely to be a biased statistical estimator for the CFR.
If we assume that the reporting of deaths is error-free, and that the true number of recoveries is $R(t) = \xi r(t)$, then $CFR = \frac{D(t)}{D(t)+R(t)} = \frac{\rho}{\rho+\xi} \quad \Rightarrow \quad \xi = \rho(\frac{1-CFR}{CFR})$

If we assume a CFR of 0.023 (2.3%) [2], we can use the reported data for $\rho$ to estimate the testing fraction $\xi$. We do this analysis for Kerala only, since KL occupies the top spot in the country on the Health Index and hence may be assumed to have accurate mortality reporting. This analysis is meaningful only in the initial stages of the epidemic, and hence we plot this ratio $\rho$ for one week after the first reported death in KL.

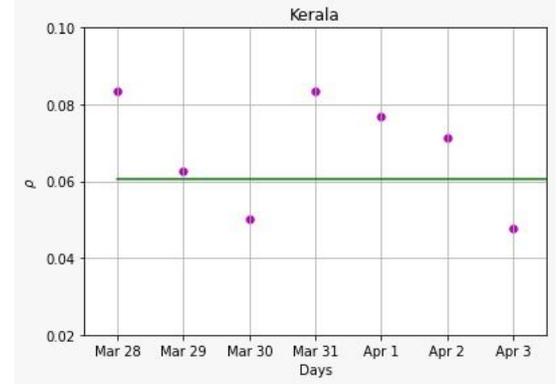

Assuming a mean value of $\rho_{mean} = 0.06$, we obtain, $\xi = 0.06 \left(\frac{1-0.023}{0.023}\right) \approx 2.5$
This means that for every 2.5 infected people in Kerala, one is identified by the testing regimen.
The testing rate for Kerala is then $r_{test} = \frac{1}{2.5} \approx 0.4$

We now use this testing fraction for Kerala to obtain estimates for testing fractions of all other states, assuming testing performances for different states correlate with states performances on two metrics - (i) the Health preparedness score of a particular state according to the National Health Mission [3], and (ii) the number of Covid-19 tests per million population for a state [4]. We scale the testing rate for an individual state according to the following empirical formula:

$$r_{test}^{STATE} = r_{test}^{KL} \times \frac{TPM_{state}}{TPM_{KL}} \times \frac{HI_{state}}{HI_{KL}}$$

We note that while testing rates is not obtained from the data for each state, it builds in a heterogeneity in the relative response of states based on the long-term health indices, and the number of covid tests conducted by each state, and hence can be assumed to be a accurate metric of the relative performance of different states. For the district level simulation, we assume all the districts for a particular state have the same testing rate.

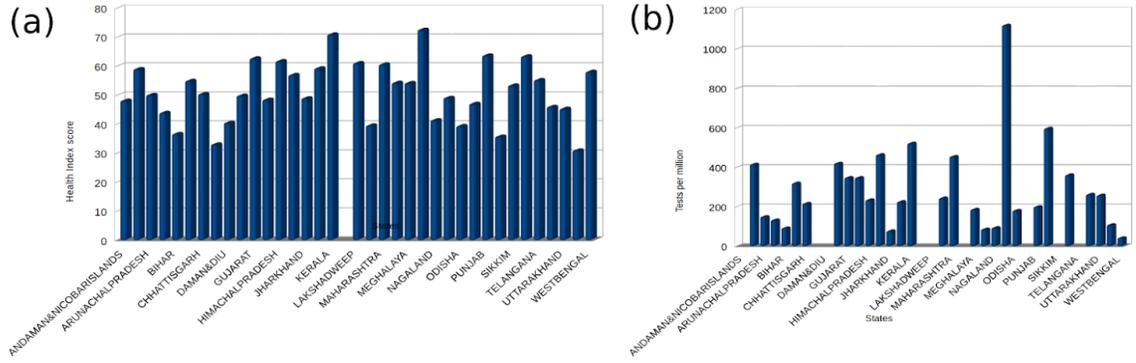

Fig. : (a) The Health Index values and (b) the number of covid tests per million conducted by individual states, used to determine testing fractions.

| State/UT | | 1 | $r_{test}$ |
|---|---|---|---|
| ANDAMAN & NICOBAR IS. | 0.029 | 0.56 | 0.18 |
| ANDHRAPRADESH | 0.03 | 0.45 | 0.265 |
| ARUNACHALPRADESH | 0.029 | 0.56 | 0.081 |
| ASSAM | 0.03 | 0.56 | 0.062 |
| BIHAR | 0.03 | 0.15 | 0.036 |
| CHANDIGARH | 0.015 | 0.65 | 0.190 |
| CHHATTISGARH | 0.025 | 0.6 | 0.118 |
| DADRA & NAGAR HAVELI | 0.029 | 0.56 | 0.036 |
| DAMAN & DIU | 0.029 | 0.56 | 0.077 |
| GOA | 0.029 | 0.56 | 0.226 |
| GUJARAT | 0.025 | 0.95 | 0.235 |
| HARYANA | 0.005 | 0.56 | 0.182 |
| HIMACHALPRADESH | 0.01 | 0.56 | 0.156 |
| JAMMU & KASHMIR | 0.02 | 0.95 | 0.286 |
| JHARKHAND | 0.03 | 0.05 | 0.041 |
| KARNATAKA | 0.015 | 0.75 | 0.144 |
| KERALA | 0.015 | 0.56 | 0.400 |
| LADAKH | 0.029 | 0.56 | 0.285 |
| LAKSHADWEEP | 0.029 | 0.56 | 0.15 |
| MADHYAPRADESH | 0.03 | 0.55 | 0.104 |
| MAHARASHTRA | 0.02 | 0.9 | 0.298 |
| MANIPUR | 0.029 | 0.56 | 0.109 |
| MEGHALAYA | 0.03 | 0.56 | 0.109 |
| MIZORAM | 0.029 | 0.56 | 0.067 |
| NAGALAND | 0.029 | 0.56 | 0.042 |
| NCT OF DELHI | 0.03 | 0.95 | 0.594 |
| ODISHA | 0.02 | 0.4 | 0.077 |
| PUDUCHERRY | 0.029 | 0.56 | 0.13 |
| PUNJAB | 0.025 | 0.25 | 0.137 |
| RAJASTHAN | 0.03 | 0.6 | 0.230 |
| SIKKIM | 0.029 | 0.56 | 0.109 |
| TAMILNADU | 0.03 | 0.56 | 0.247 |
| TELANGANA | 0.03 | 0.3 | 0.109 |
| TRIPURA | 0.029 | 0.56 | 0.130 |
| UTTARAKHAND | 0.015 | 0.65 | 0.126 |
| UTTARPRADESH | 0.015 | 0.95 | 0.036 |
| WESTBENGAL | 0.03 | 0.05 | 0.026 |

**Mortality rates:**

In order to calculate the mortality rates, we first determine the population distribution of each state into ten-year bins. The mortality rates of each of these 10-year bins are taken from the literature [55] and are listed in the table below. We then appropriately weigh the population distribution of each state in order to determine a state-specific mortality rate for each of the three age groups in our model (0-20, 20-60, >60).

| Age group | 0-9 | 10-19 | 20-29 | 30-39 | 40-49 | 50-59 | 60-69 | 70-79 | >80 |
|---|---|---|---|---|---|---|---|---|---|
| Mortality rate | 0.0 | 0.2 | 0.2 | 0.2 | 0.4 | 1.3 | 3.6 | 8.0 | 14.8 |

## C. Equilibrium and model reproduction number

We will consider the disease compartments of the model described above without the lockdown compartments to calculate the basic reproduction rate using the next generation matrix method. The evolution of the disease compartments are given by

$$\frac{dI}{dt} = \sigma A - \kappa_t I - \gamma_2 I,$$
$$\frac{dA}{dt} = \bar{f}\beta(I + A + b_3 P + \beta_2 E)S/N - \sigma A - \kappa_t'' A,$$
$$\frac{dP}{dt} = \kappa_t I + \kappa_t' E + \kappa_t'' A - \gamma_3 P,$$
$$\frac{dE}{dt} = f\beta(I + A + b_3 P + \beta_2 E)S/N - \gamma_1 E - \kappa_t' E.$$

The usual procedure follows by evaluating the Jacobian of the sub-model evolution separated into three parts. This yields the two matrices

$$F = \begin{bmatrix} 0 & 0 & 0 & 0 \\ \bar{f}\beta S(t)/N & \bar{f}\beta S(t)/N & \bar{f}\beta b_3 S(t)/N & \bar{f}\beta\beta_2 S(t)/N \\ 0 & 0 & 0 & 0 \\ f\beta S(t)/N & f\beta S(t)/N & f\beta b_3 S(t)/N & f\beta\beta_2 S(t)/N \end{bmatrix},$$

and

$$V = \begin{bmatrix} \kappa_t + \gamma_2 & -\sigma & 0 & 0 \\ 0 & \sigma + \kappa_t'' & 0 & 0 \\ -\kappa_t & -\kappa_t'' & \gamma_3 & -\kappa_t' \\ 0 & 0 & 0 & \gamma_1 + \kappa_t' \end{bmatrix}.$$

Basic Reproduction Number $R_0$ = Spectral Radius ($FV^{-1}$)

$$R_0 = \underbrace{\frac{f\beta\beta_2}{\gamma_1 + \kappa_t'}}_{\text{from E}} + \underbrace{\frac{\bar{f}\beta}{\sigma + \kappa_t''}}_{\text{from A}} + \underbrace{\frac{\bar{f}\beta\sigma}{(\gamma_2 + \kappa_t)(\sigma + \kappa_t'')}}_{\text{from I}} +$$

$$\underbrace{\frac{\bar{f}\beta b_3 \sigma \kappa_t}{\gamma_3(\gamma_2 + \kappa_t)(\sigma + \kappa_t'')} + \frac{\bar{f}\beta b_3 \kappa_t''}{\gamma_3(\sigma + \kappa_t'')} + \frac{f\beta\kappa_t' b_3}{\gamma_3(\gamma_1 + \kappa_t')}}_{\text{from P}}$$

For completeness, we note two limits of the evaluated $R_0$.

**Reduction to SEIR model for $f = 1$ and $\kappa_t' = 0$:** This model reduces to the basic SIR model if we have $f = 1$ and $\kappa_t' = 0$. E bin of this model is the I bin of the SIR model. For such a case:

$$R_0 = \beta\beta_2/\gamma_1,$$

where $\beta_2$ is the infection rate and $\gamma_1$ is the recovery rate.

**Reduction to SEIR model for $f = 0$, $\kappa_t' = 0$, $\kappa_t'' = 0$, and $b_3 = 0$:** For these given values of parameters our model reduces to the SEIR model and A bin of our model is the E bin of SEIR model. Buth the basic difference in the SEIR model and our special case is that the infection rate for their model is $(I) = I/N$ while for our case it becomes $(I, A) = (I + A)/N$. Hence $R_0$ for our case is slightly different from the original SEIR model and is given as:

$$R_0 = \beta\left(\frac{1}{\sigma} + \frac{1}{\gamma_2}\right).$$

### D. Extended Kalman Filter

For a given model, the process of estimating the true state of the model, given the equations for state dynamics and measurements, is a standard problem in estimation theory. The state vector of our system can be written as $\psi = (S, X_S, A, X_A, E, X_E, I, X_I, P, R)$ for the fundamental unit cell of our model (which we use to outline the estimation theory). The problem can now be stated as estimating $\psi_i \in R$, $\psi_i \geq 0$ given the two noisy measurements namely $P(t_i)$ and mortality $m(t_i)$ for each day $t_i$.

The state equations for a generic state vector $x$ for a nonlinear state space model, can be expressed as:

$$\dot{x}_k = f(x_k, u_k, \eta_k),$$
$$y_k = h(x_k, v_k),$$

where $u_k$ represents the control input to the system. $\eta_k$ and $v_k$ represent the process noise and measurement noise respectively.

For our system model, several modelling assumptions can be made that make the system amenable for Kalman filtering while preserving its stochastic properties. The evolution of $\psi$ is not governed by any control inputs. We assume that $\eta_k$ and $v_k$ are additive noises to the respective equations. The state transition equation is further discretized and the cumulative effect of the process noises is incorporated within an additive noise $w_k$. Hence, the state equations of the discrete-time nonlinear system described by $\psi$ is:

$$\psi_{k+1} = F(\psi_k) + w_k,$$
$$z_k = h(\psi_k) + v_k,$$

where

$$F(\psi_k) = \psi_k + \int_{t_k}^{t_{k+1}} f(\psi(\tau))\, d(\tau).$$

The disturbances $w_k$ and $v_k$ are assumed to be Gaussian, that is,

$$w_k \in \mathcal{N}(0, \Sigma_Q), \quad \text{and} \quad v_k \in \mathcal{N}(0, \Sigma_R).$$

The function $f(\psi(\tau))$, $\tau \in \{t_k, t_{k+1}\}$ is obtained from the time evolutions of each compartment provided earlier in Section A. We obtain $F(\psi_k)$ by numerically integrating the differential equation in real time.

The measurements are composed of the compartment that tested positive as well as the mortality. The latter is calculated heuristically based on the compartments of symptomatic people, and on the mortality rate which depends on the age bin and the state. Hence $h(\psi(\tau))$ can be expressed as:

$$h(\psi_k) = [P_k \quad m_{state, bin}(I_k + X_{Ik} + P_k)]^{\mathrm{T}}$$

We define the observer $\hat{\psi}_{k|k}$, which is the estimate of the state vector $\psi_k$ at time instant $t_k$, with the following state dynamics and measurement model:

$$\begin{aligned}
\hat{\psi}_{k+1|k} &= F(\hat{\psi}_{k|k}) \approx \phi_k \hat{\psi}_{k|k}, \\
\hat{z}_{k+1|k} &= h(\hat{\psi}_{k+1|k}) \approx H_k \hat{\psi}_{k+1|k}.
\end{aligned}$$

Where,

$$\phi_k = \frac{\partial F(\hat{\psi}_{k|k})}{\partial \hat{\psi}_{k|k}}, \text{ and } H_{k+1} = \frac{\partial h(\hat{\psi}_{k+1|k})}{\partial \hat{\psi}_{k+1|k}}.$$

The observer state vector $\hat{\psi}_{k|k}$ is a multivariate random variable, as is the case with every kind of optimal estimator. For this system, it is of multivariate Gaussian form and has a covariance matrix $\Sigma_{k|k}$.

The prediction step of the Extended Kalman Filter (henceforth referred to as the EKF) consists of the following steps:

$$\begin{aligned}
\hat{\psi}_{k+1|k} &= F(\hat{\psi}_{k|k}), \\
\Sigma_{k+1|k} &= \phi_k \Sigma_{k|k} \phi_k^{\mathrm{T}} + \Sigma_Q.
\end{aligned}$$

This prediction is updated based on the measurements obtained. The equations for the update step are as follows:

$$\begin{aligned}
\hat{z}_{k+1|k} &= h(\hat{\psi}_{k+1|k}), \\
V_{k+1} &= H_{k+1} \Sigma_{k+1|k} H_{k+1}^{\mathrm{T}} + \Sigma_R, \\
L_{k+1} &= \Sigma_{k+1|k} H_{k+1}^{\mathrm{T}} V_{k+1}^{-1}, \\
\hat{\psi}_{k+1|k+1} &= \hat{\psi}_{k+1|k} + L_{k+1}(z_{k+1} - \hat{z}_{k+1|k}), \\
\Sigma_{k+1|k+1} &= (I - L_{k+1} H_{k+1}) \Sigma_{k+1|k} (I - L_{k+1} H_{k+1})^{\mathrm{T}} + L_{k+1} \Sigma_R L_{k+1}^{\mathrm{T}}.
\end{aligned}$$

The dataset has measurements upto 3rd May, hence for further dates, we use the predictions in the following manner:

$$\psi_{k+1|k+1} = \psi_{k+1|k} \text{ and } \Sigma_{k+1|k+1} = \Sigma_{k+1|k}.$$

### E. Transport matrix

The process to develop the inter-district transportation matrix representing the population commuting between adjoining districts is shown in Figure D1. For this, the trip length frequency distribution, obtained from the worker population trip length recorded during the 2011 census data collection, was used to estimate the probability density function of trip length. Worker population ($WP$) residing at $DDB$ distance from the district boundary may either commute towards the district boundary or in directions not leading to the district boundary. We assumed half of them would travel toward the district boundary. When commuting towards the district boundary, workers would cross the boundary if their trip length ($TL$) exceeds the $DDB$. Hence, workers expected to commute across the district boundary can be obtained from $\frac{WP \times P(TL>DDB)}{2}$.

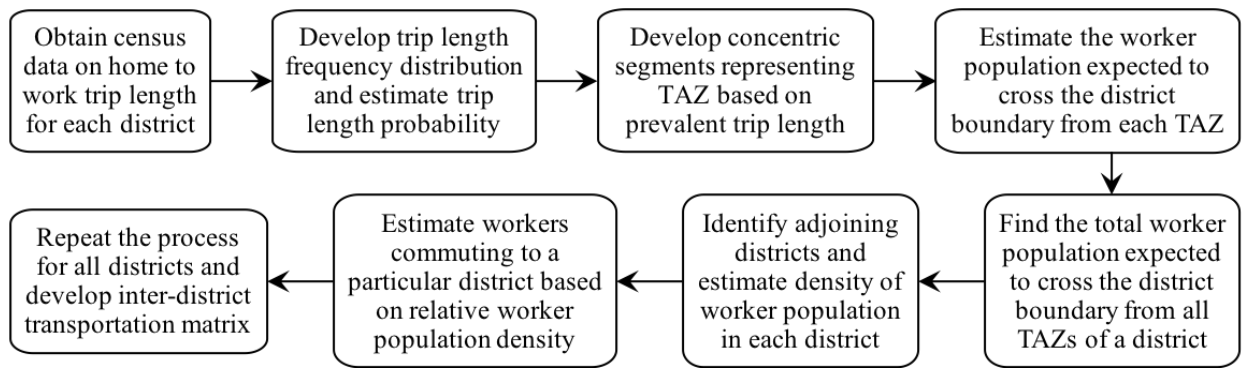

Figure D1: Flowchart to develop the inter-district transportation matrix

The GIS map of each district was segmented as shown in Figure D2 to discretize the $DDB$. Each of the segmented zones represents the traffic analysis zone (TAZ). The $DDB$ adopted for each TAZ were 1 km, 5 km, 10, km, 20, km, 30 km and 50 km and the total worker population ($TP_i$) expected to commute across the district boundary was estimated from Equation D1.

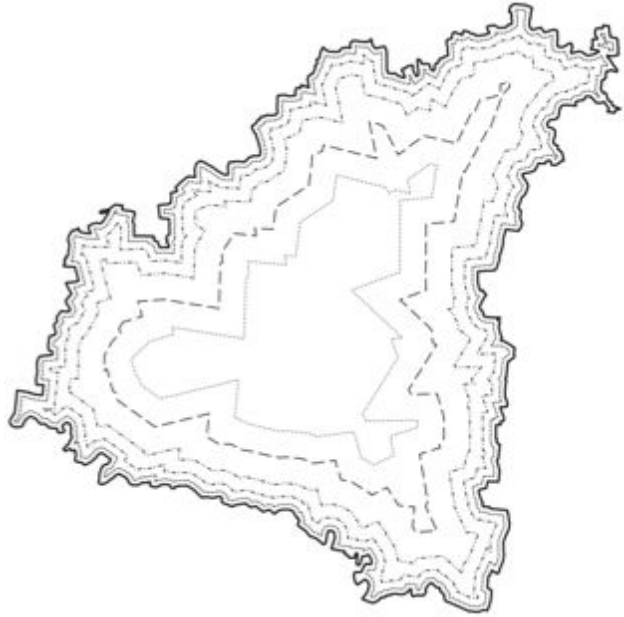

Figure D2: Segmentation of a typical district map for traffic analysis zones

$$TP_i = \frac{\sum_{S=1}^{N} WP^S \times P(TL > DDB^S_{avg})}{2} \qquad \text{D1}$$

where,

| $TP_i$ | = | Total worker population of district $i$ expected to cross the district boundary |
| --- | --- | --- |
| $WP^S$ | = | Worker population in segment $S$ |
| $DDB^S_{avg}$ | = | Average distance to district boundary from segment $S$ |
| $N$ | = | Total number of segments |

The adjoining districts would attract the commuting workers based on its economic status. A district level GDP (at purchasing power parity) per capita can be a good economic status indicator. However, in absence of that we adopted the relative worker population density as a surrogate measure to estimate the workers commuting to a particular adjoining district. Equation D2 is used for the purpose. This process is repeated for all the districts of India to come up with the inter-district transportation or mobility matrix.

$$M_{ij} = TP_i \frac{PD_j}{\sum_{j=1}^{AD} PD_j} \qquad \text{D2}$$

where,

| $M_{ij}$ | = | Worker population expected to commute from district $i$ to $j$ |
| --- | --- | --- |

| $PD_j$ | = | Worker population density of district $j$ |
|---|---|---|
| $AD$ | = | Total number of adjoining districts |